\documentclass[superscriptaddress,showpacs,twocolumn,aps,pre,10pt]{revtex4-1}
\usepackage{natbib}
\usepackage{graphicx,dcolumn,bm,bbm,amsmath,amssymb,color}

\newcommand{\eeq}{ \end{equation} }
\newcommand{\beq}{ \begin{equation} }
\newcommand{\bea}{\begin{eqnarray}}
\newcommand{\eea}{\end{eqnarray}}

\begin{document}
\title{Effect of self-propulsion on equilibrium clustering}
 
\author{Ethayaraja Mani}
\email{ethaya@iitm.ac.in}
\affiliation{Polymer Engineering and Colloid Science Lab, Department of Chemical Engineering, Indian Institute of Technology Madras, Chennai - 600036, India}
\author{Hartmut L\"{o}wen}
\affiliation{Institut f\"ur Theoretische Physik II: Weiche Materie,
Heinrich-Heine-Universit\"at D\"{u}sseldorf,
Universit{\"a}tsstra{\ss}e 1, D-40225 D\"{u}sseldorf,
Germany}
 
\date{\today}

\begin{abstract}
In equilibrium, colloidal suspensions governed by short-range attractive and long-range repulsive interactions
form thermodynamically stable clusters. Using Brownian dynamics computer simulations,
 we investigate how this equilibrium clustering is affected when such particles are self-propelled. We find that the clustering process is stable under self-propulsion. For the range of interaction parameters studied and at low particle density, the cluster size increases with the speed of self-propulsion (activity) and for higher activity the cluster size decreases, showing a non-monotonic variation of cluster size with activity. This clustering behaviour is distinct
 from the pure kinetic (or motility-induced) clustering of self-propelling particles which is observed at
significantly higher activities and densities. We present an equilibrium model incorporating the effect of activity as activity-induced attraction and repulsion by imposing that the strength of these interactions depend on activity superlinearly. The model explains the cluster size dependence of activity obtained from simulations semi-quantitatively. Our predictions are verifiable in experiments on interacting synthetic colloidal microswimmers.
\end{abstract}
 
\maketitle
\section{Introduction}
Concentrated colloidal or protein solutions which are governed by
a combination of short-range attractive and long-range repulsive interaction potentials exhibit
a stable clustering phenomenon in equilibrium at finite temperature and moderate densities. This phenomenon is first predicted by theory \cite{kegel} and simulation \cite{sciortino,imperio,godfrin,mani}
and has been confirmed in experiments \cite{stradner,bartlett}. The intuitive explanation \cite{kegel}
for equilibrium clustering
lies in the fact that the short-ranged attraction first leads to growth of clusters in an initially
dilute suspensions of particles. The growth stops, however,
 when the cluster size reaches a characteristic size where the
long-ranged repulsion leads to an increase in the  self-energy of the cluster. In equilibrium,
at finite temperature, this leads to a typical average cluster size which depends
on the interaction parameters and the imposed global particle density.
While the details of this equilibrium cluster process are understood for a decade by now,
recent developments have considered self-propelled  (or active) particles which dissipate energy
leading to synthetic microswimmers \cite{schimansky,marchetti}.
These particles also exhibit a purely kinetic clustering
if the strength of self-propulsion is sufficiently large \cite{marchetti,review}
which has recently been found in experiments
\cite{berthier,buttinoni,chaikin} and explored by simulation
 \cite{buttinoni, fily2012, stenhammar2013, redner, bialkeepl2013, stenhammar2014, fily2014, speck, wysocki,mognetti} and theory
\cite{tailleur,cates2010,cates2013,bialkeepl2013, speck,wittkowski,peruani,cremer}.
This purely motility-induced clustering occurs for repulsive systems and
is therefore absent in equilibrium (i.e.\ for
vanishing drive). The study of Redner et al \cite{redner2013} showed reentrant phase behavior in active Lennard-Jones particles, where attractive interaction and activity compete to stabilize phase-separated states at low and high activities, respectively.  

Here we link the fields of equilibrium clustering to that of microswimmers.
We consider the equilibrium clustering and study how this is affected by an imposed self-propulsion.
The motivation to do so is  three-fold. First of all, this is an interesting problem in itself since upon
increasing the self-propulsion, there are two counterbalancing effects: on the one hand,
the self-propulsion leads to a
higher mobility and hence an effect which is expected to correspond to an increase of temperature.
On the other hand, however, the self-propulsion yields a larger sticking probability
of neighbouring particles which
would favor and enhance the clustering tendency. The second motivation comes from the fact that
one needs to understand whether there is a hidden pathway between the two different kinds of clustering
mentioned above, i.e.\ to check whether they are distinct or interconnected in a certain parameter space.
Finally, artificial colloidal model microswimmers can be prepared with controlled interactions, e.g. by
adding depletants \cite{poonpnas}, tuned van der Waals attractions
or charging the particles such that  model colloidal
swimmers can be prepared, in principle, with short-ranged attraction and long-ranged repulsion. The additional
tunability of the interparticle potential then allows to control the degree of clustering
which needs a systematic understanding.

In this paper
we simulate a two-dimensional model of microswimmers with competing interactions by using Brownian
dynamics computer simulations.
We use a model proposed by Sear et al \cite{sear} and Imperio and Reatto \cite{imperio}, for which the equilibrium clustering behaviour
is well-understood in two-dimensions but supplement this here for an additional self-propulsion
in the simplest form by neglecting explicit alignment and hydrodynamic interactions \cite{bialkeprl2012}.
As a result, we find indeed that the trends of clustering depend on the interaction parameters.
The self-propulsion can either increase or decrease the cluster size. In fact there is a complex
and maybe unanticipated non-monotonic behaviour of the cluster size as a function of increasing
self-propulsion: it can first increase and then decrease again. This cannot be understood by simple
temperature rescaling as has been previously noted for active systems
in the context of freezing \cite{bialkeprl2012} as well as  in a trapping \cite{szamel} and  in a
gravitational field \cite{berthier}. Dynamic (or purely motility-induced) clustering also occurs in our model although at
much larger drives, where the details of the interactions become irrelevant. In this case, the cluster sizes are rather small compared to that of equilibrium clusters. Thus the two clustering phenomena appear quite distinct.
\section{Model and Simulation}
\begin{figure} 
\centering
\includegraphics [scale=0.37]{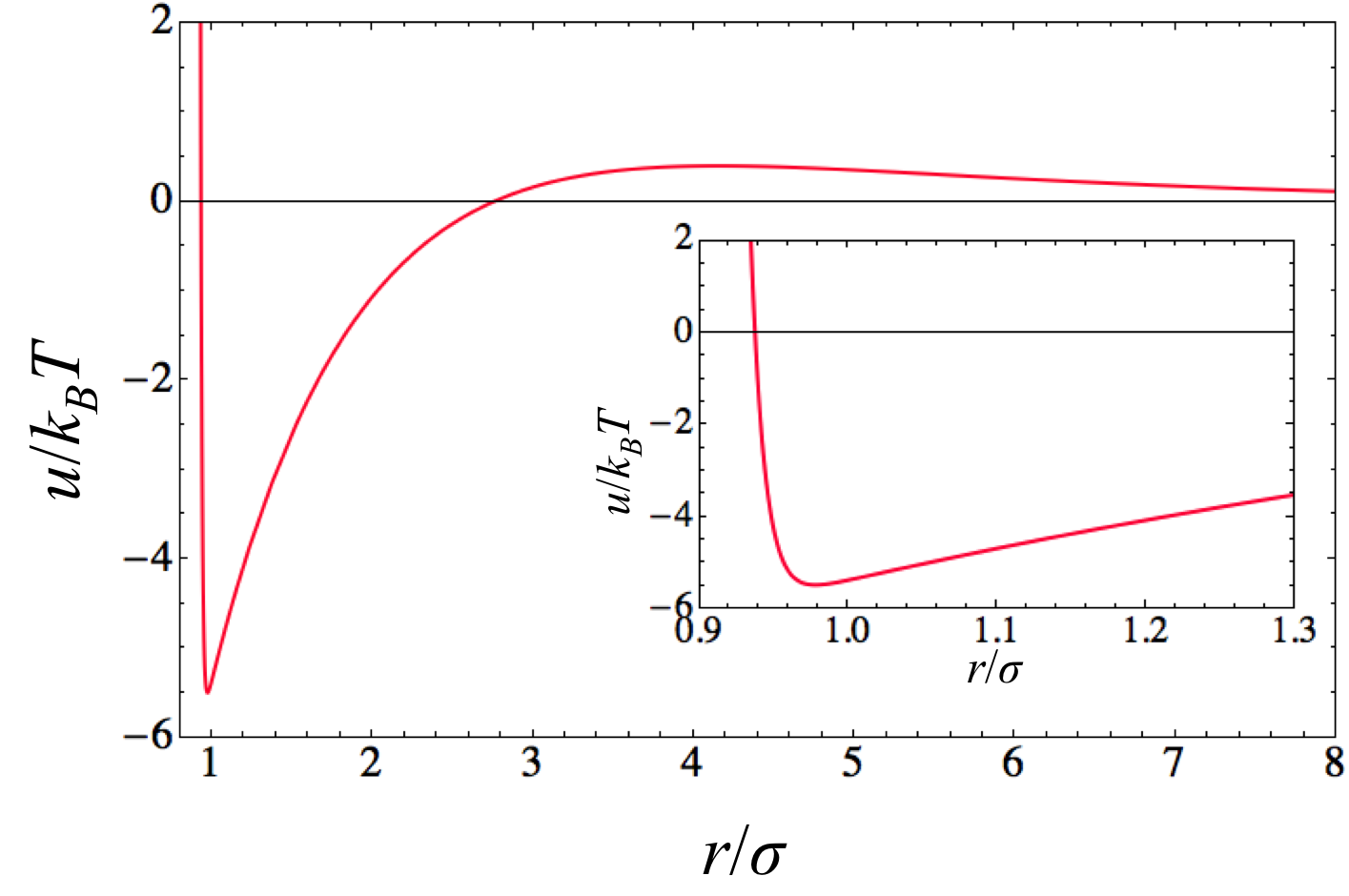}
\caption{ (Colour online) Overall interaction potential $u(r)$ showing competing attractive and repulsive interactions for ${\epsilon_a} = 25 k_B T$ from Eq. \ref{eq:1}. The inset shows the potential near contact.} 
\label {fig:potential}
\end{figure}

We model short-range attractive and long-range repulsive  interactions
using a modified Lennard-Jones potential $u_{1}(r)$ and a double-exponential potential $ u_{2}(r)$
that was introduced previously to explain the formation of finite sized clusters and stripes of nanoparticles at air-water interface \cite{sear}. The overall potential is given as:
\begin{equation} \label {eq:1} 
\begin{aligned}
u(r) = u_{1}(r) + u_{2}(r)
\end{aligned} 
\end{equation}
Here $u_{1}(r)$ is defined as
\begin{equation} \label {eq:2} 
\begin{aligned}
u_{1}(r) = 4\epsilon_{LJ} \left[ \left( \frac {\sigma_{LJ}} r \right)^{100} -\left( \frac {\sigma_{LJ}} r \right)^{50} \right]
\end{aligned} 
\end{equation}
and $u_{2}(r)$ is given by
\begin{equation} \label {eq:3} 
\begin{aligned}
u_{2}(r) = -\frac {\epsilon_a \sigma^2} {R_a^2}  \exp \left(-\frac r R_a \right) + \frac {\epsilon_r \sigma^2} {R_r^2}  \exp \left(-\frac r R_r \right) 
\end{aligned} 
\end{equation}
Here \textit{r} is the interparticle distance, and ${\sigma}$ is the diameter of the particle. ${\epsilon_{LJ}}$ and ${\sigma_{LJ}}$ are the parameters of 
the modified Lennard-Jones potential. We fix the potential parameters to $\sigma_{LJ}$ = $\sigma$, $R_{a}$ = ${\sigma}$, $R_{r}$ = 2${\sigma}$ and ${\epsilon_a}$ = ${\epsilon_r}$. In  the following, we use dimensionless quantities and express energy in units of $k_B T$,
length in units of ${\sigma}$, time in units of $\tau = {\sigma}^2/D$. Here, $k_B$ is Boltzmann constant, $T$ is temperature and $D$ is the diffusion coefficient of a single passive particle. We  further fix $\epsilon_{LJ} = 0.0025k_B T$. Fig.\ref{fig:potential} shows the variation of potential with interparticle distance for ${\epsilon_a} = 25k_B T$. Note that the repulsive part of the double-exponential potential is rather long ranged.  

Brownian dynamics simulations in two dimensions in the $xy$-plane
are performed with particles interacting via the potential given in Eq.\eqref{eq:1}.
We simulate $N = 1050$ particles using a square box with periodic boundary conditions.
To mimic self-propulsion, the particles are defined with an orientation $\textbf{e}_i$ diffusing freely about
the perpendicular \textit{z} axis with rotational diffusivity  $D_r$. In two dimensions, the components of $\textbf{e}_i$  are given as
 $\textbf{e}_i = (\cos \varphi_i,  \sin \varphi_i)$.
 In addition to translational Brownian motion,
the particles are driven with constant speed \textit{v} along their  orientation $\textbf{e}_i$.
Here there are no aligning interactions as the pair potential is independent of orientations.
Moreover hydrodynamics interactions are neglected. The resulting equations of motion for the
particle positions \{$\textbf{r}_i$\} and orientations \{$\textbf{e}_i$\} are then given by
\begin{equation} \label {eq:3}
\begin{aligned}
{\dot {\textbf{r}}}_i = \frac {D} {k_{B}T}(-\nabla_{\textbf{r}_i} U) + \textbf{e}_iv + \boldsymbol{\xi}_i
 \end{aligned} 
\end{equation} 
\begin{equation} \label {eq:torque}
\begin{aligned}
{\dot {{\varphi}}}_i =  {\xi}_i^r
 \end{aligned} 
\end{equation}  
Here, $U = \sum\nolimits_{i<j} u(|\textbf{r}_i -\textbf{r}_{j}|)$ is the total pair potential. The self-propulsion
speed of the particle is referred to in terms of the dimensionless Peclet number $Pe$ defined as
\begin{equation} \label {eq:pe}
\begin{aligned}
Pe = \frac {\sigma v} {D}
\end{aligned}
\end{equation}
The Gaussian noise $\boldsymbol{\xi}_i$ models the stochastic  solvent kicks. It has a zero mean and variance
$\langle \boldsymbol{\xi}_i(t)\boldsymbol{\xi}_j^T(t^\prime)\rangle = 2\delta_{ij}D\textbf{1}\delta(t-t^\prime)$, where \textbf{1} is the identity matrix. Similarly, the stochastic random torque
 ${\xi}_i^r$ has a zero mean and a variance of $\langle {\xi}_i^r(t){\xi}_j^r(t^\prime)\rangle = 2D_r\delta_{ij}\delta(t-t^\prime)$.
The rotational diffusivity is taken as $D_r = 3{D}/\sigma^2$, which is a valid approximation for a spherical
particle undergoing free rotational diffusion.  
The equations of motion Eq. \eqref{eq:3} and \eqref{eq:torque} are numerically integrated
with a time step of $10^{-5}\tau$. The long range potential Eq.\eqref{eq:1}
is truncated at $r = 15\sigma$. Simulations were done for  various reduced areal densities (typically of $0.13\sigma^{-2}$ unless stated otherwise) and for different values of ${\epsilon_a}$. The simulations
are performed starting from a random initial configuration of particles. Typically the system is simulated for $500\tau$
(i.e.\ $5\times10^7$ steps) to attain a steady-state, followed by  production runs of another $500\tau$. 
The simulations are replicated 5 times with different initial random configurations and the properties 
are time-averaged over all the replicas. Careful tests were performed to check that the system achieved a steady state by monitoring the saturation of the cluster size as a function of time and by making sure that there is enough exchange dynamics between the clusters.
 
 To define a cluster we use a cutoff distance
 of 1.5${\sigma}$, which corresponds to the interparticle distance where the potential is roughly half of the potential at contact.
 Two particles belong to the same cluster if they are connected by a sequence of other particles which are all
 separated by less than 1.5${\sigma}$.
 The average
 cluster size is calculated from
\begin{equation} \label {eq:cluster}
\begin{aligned}
<\textit{n}> =\sum \limits_{n=1}^N n P(n) 
 \end{aligned}
\end{equation}
Here, $P(n)$ is the probability to find a cluster of \textit{n} number of particles at steady state. We also monitor the fluctuations in the cluster size by calculating the reduced variance
\begin{equation} \label {eq:var}
\begin{aligned}
Var(n) =\frac {<n^2> - <n>^2} {<n>^2}.
 \end{aligned}
\end{equation}

\section{Clustering of active particles}
\begin{figure} 
\centering
\includegraphics [scale=0.37]{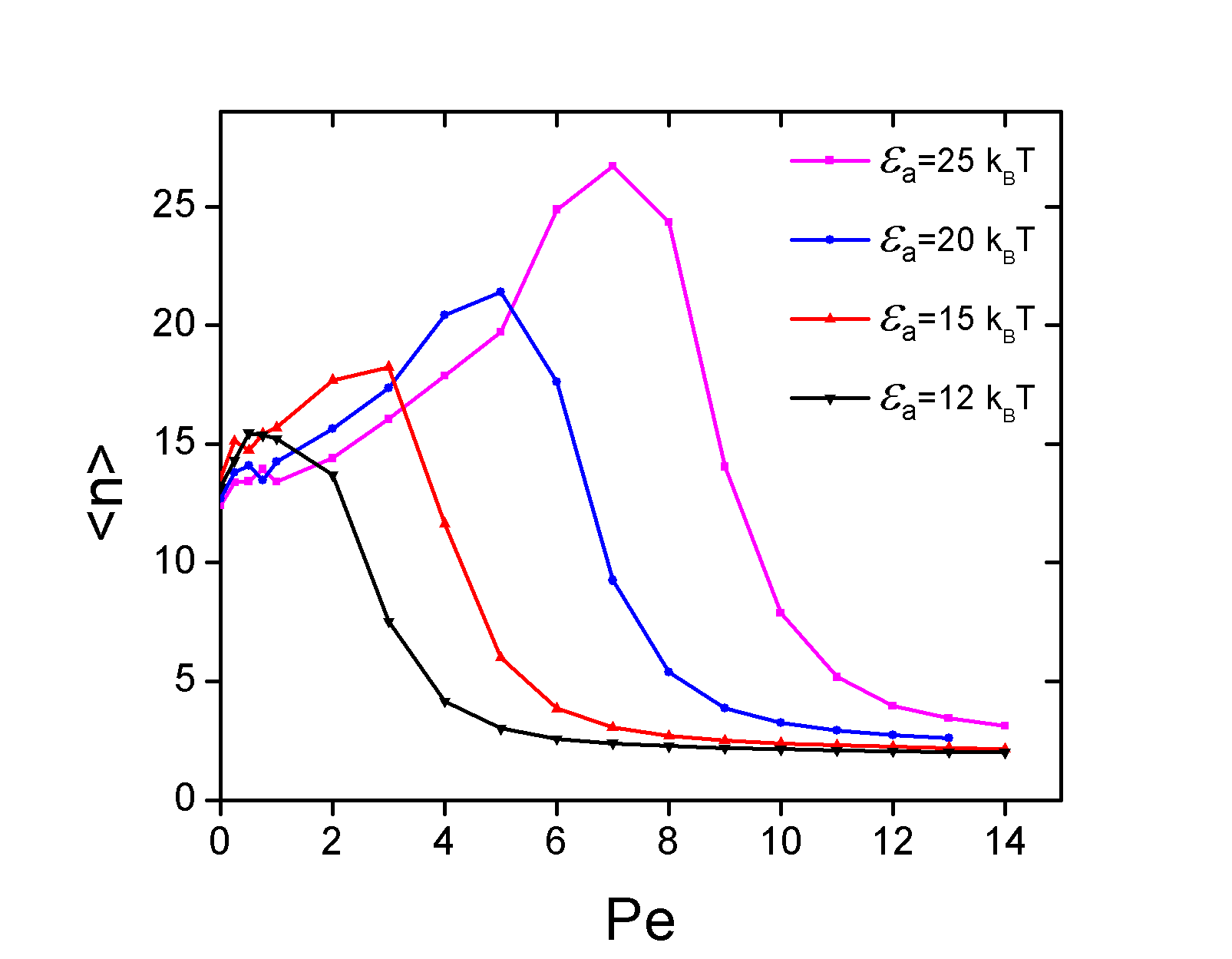}
\caption{ (Colour online) Effect of activity on the average cluster size $<n>$ for various values of ${\epsilon_a}$.}
\label {fig:activity}
\end{figure}

Passive particles with short-ranged attractive and long-ranged repulsive interactions defined in Eq. \eqref{eq:1} show equilibrium clustering at certain densities. In particular, equilibrium clustering occurs for a density of about 
0.13 and for attraction energies ${\epsilon_a}$ in the range of $12k_B T$ to $25 k_B T$ \cite{imperio}. These clusters are quasi-circular in shape. The effect of activity on clustering is shown in Fig. \ref{fig:activity} where the  average cluster size $<n>$ is shown versus the Peclet number $Pe$. For the cases of vanishing activity, the average cluster size is about 
$<n> \approx 14$ independent of ${\epsilon_a}$. 

\begin{figure*}
\begin{center}
\begin{tabular}{ccc}
\includegraphics[scale=0.52]{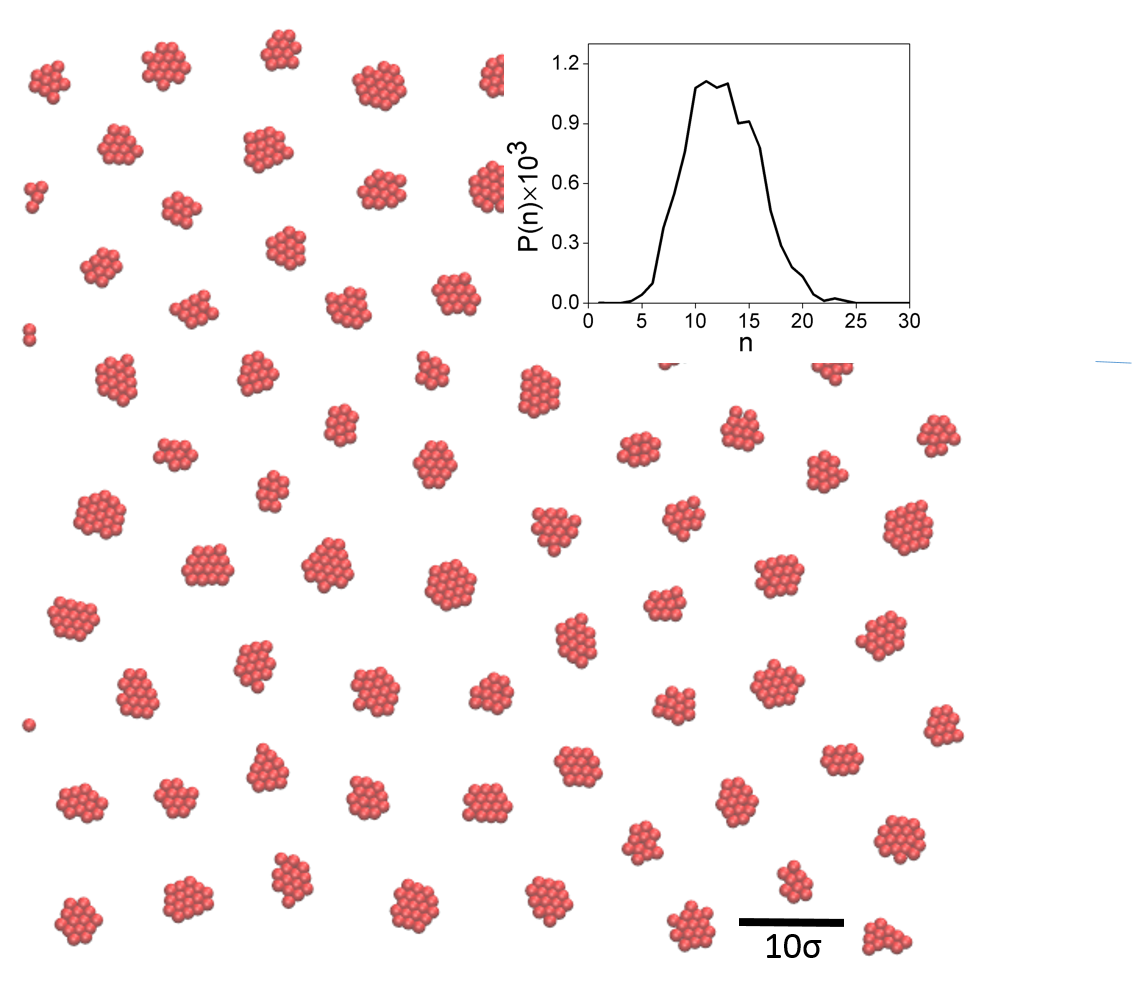}&\includegraphics[scale=0.5]{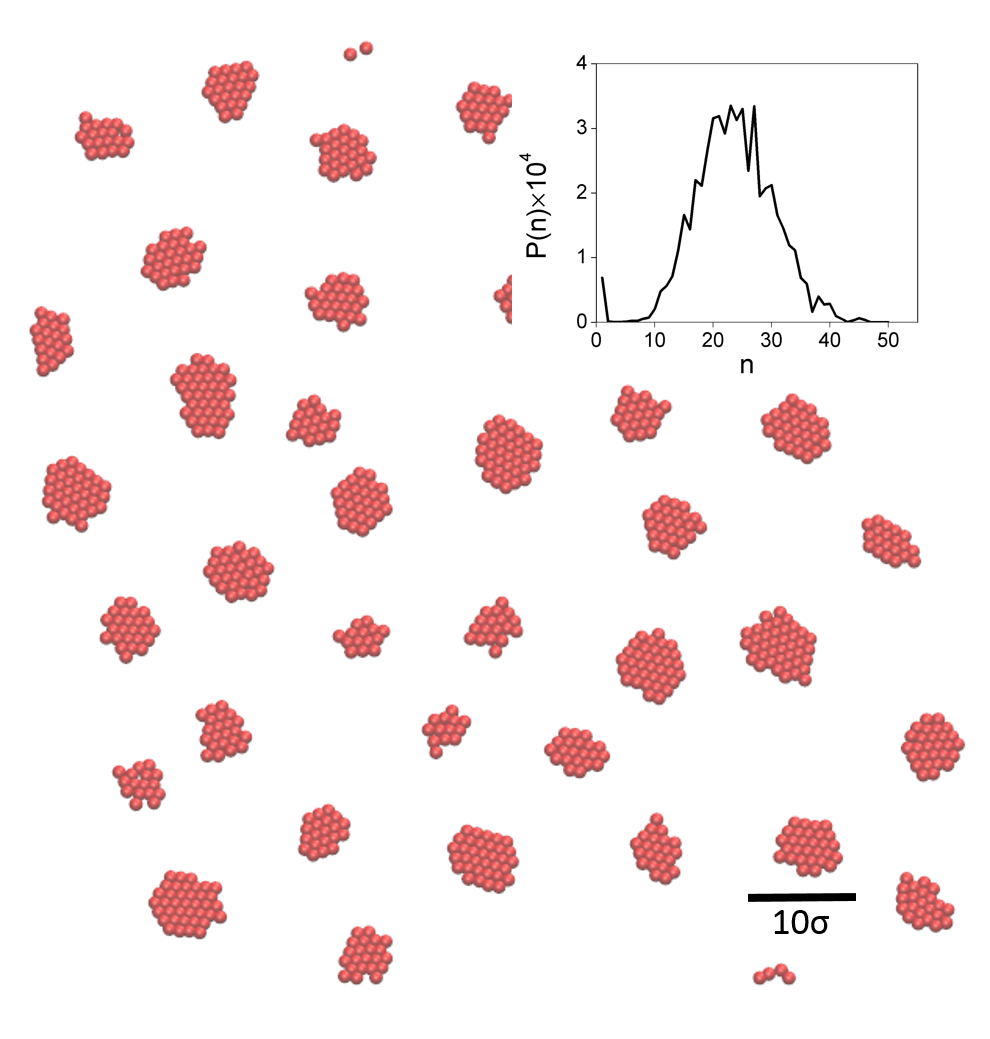}\\
(a)&(b)\\
\end{tabular}
\caption{(Colour online) Representative snapshots of clusters  for $Pe$ = 0 (a) and $Pe$ = 6 (b) for ${\epsilon_a}$ = 25 $k_BT$. The insets show the cluster size distribution $P(n)$ in the steady state.}
\label {fig:pe0}
\end{center}
\end{figure*}

Strikingly, the dependence of the average cluster size on activity is non-monotonic. The cluster size first increases with increasing activity and attains a maximum before it finally decreases at higher activity. This trend is seen for all the ${\epsilon_a}$ values studied here. The critical activity corresponding to the maximum of cluster size increases with increasing ${\epsilon_a}$. This behaviour points to the possibility that the activity manifests itself as an effective attraction which increases the cluster size until the activity gets so high that particles are eventually removed from the cluster overcoming attractive interactions between the particles in the cluster. It is interesting that we find a maximum in the cluster size at intermediate $Pe$ while Redner et al \cite{redner2013} find a suppression of phase separation at intermediate $Pe$. Therefore, our findings are qualitatively different from that of Redner et al \cite{redner2013} due to the combination of attraction and repulsion. 
  
\begin{figure} 
\centering
\includegraphics [scale=0.37]{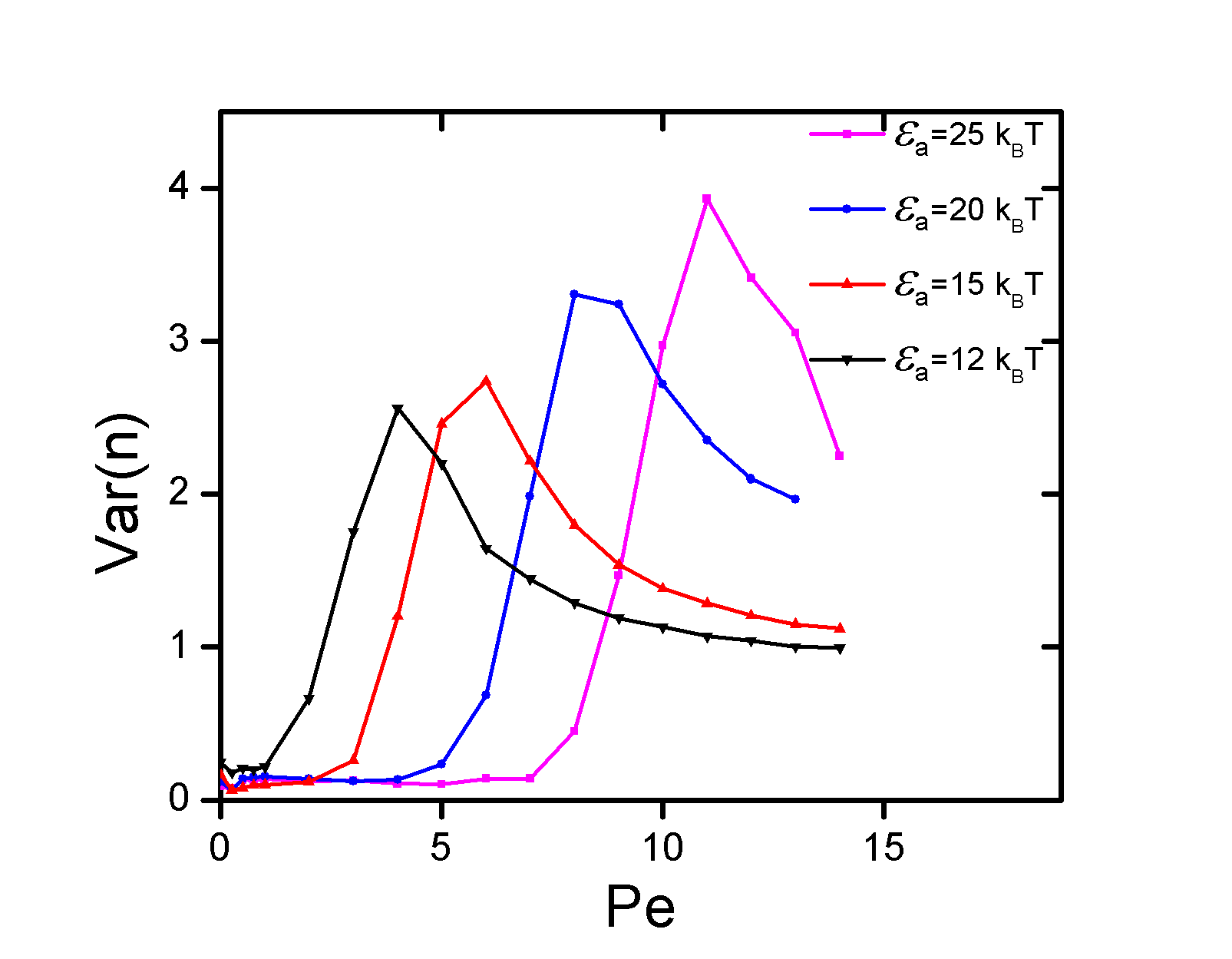}
\caption{(Colour online) Effect of activity on the reduced variance of cluster size for various ${\epsilon_a}$ values.}
\label {fig:variance}
\end{figure}

Representative snapshots showing clusters for $Pe=0$ and $Pe=6$ (at fixed $\epsilon_a =25 k_BT$) are shown in Fig. \ref{fig:pe0}a and \ref{fig:pe0}b, respectively. The clusters exhibit an inner crystalline structure as found in previous simulations for dynamical clustering \cite{buttinoni}. Moreover, there is a large spread in cluster sizes. This is also documented by  the normalized size distribution function $P(n)$ which is shown in the insets of Fig. \ref{fig:pe0}a and \ref{fig:pe0}b. An increased activity leads to a much larger distribution in the cluster size at steady-state. This is documented by the reduced variance of the cluster size distribution which steeply increases with $Pe$ as shown in Fig. \ref{fig:variance}, until it reaches a maximum and decreases as it is correlated with the average cluster size.

\section{Effective equilibrium model}
We now present a phenomenological model to explain the clustering behavior observed in active particles by balancing interparticle interactions with activity. First we consider passive particles with short-ranged attractive and long-ranged repulsive
 interactions and include the effect of activity in the model later. Consider monomeric discs of diameter $\sigma$ assembled as circular clusters of uniform diameter $d$ with $n$ number of monomers. The discs interact with themselves via a short-range attraction and long-range repulsion.  Following the approach of Groenewold and Kegel \cite{kegel}, the free energy of such a cluster of passive particles can be written as 
\begin{equation} \label {eq:m1}
\begin{aligned}
f(n) = \frac {gn^2E_r \sigma} d - nE_a + \pi d \lambda
\end{aligned}
\end{equation}
where $g$ is a geometric parameter related to circular shape,  $E_r$ and $E_a$ are typical repulsive and attractive energies of a particle inside the cluster. The first term accounts for total repulsive energy and is of order $n^2$ due to long-range nature of repulsion, the second term accounts for attractive energy assuming the attraction to be short ranged and therefore scales linear in $n$. The last term is energy due to line tension $\lambda$ of the cluster boundary. Also note that $n$ and $d$ are related via,
\begin{equation} \label {eq:m2}
\begin{aligned}
n = \frac {\pi d^2} {4a}
\end{aligned}
\end{equation}
where $a$ is the cross sectional area of the monomer ($a = \pi \sigma^2/4$). Combining Eqs. \eqref{eq:m1} and \eqref{eq:m2} we get
\begin{equation} \label {eq:m3}
\begin{aligned}
\frac {f(n)} n = \sqrt \frac {\pi}{a}gE_r \sqrt n - E_a + 2\lambda \sqrt {\pi a}\frac {1}{\sqrt n}
\end{aligned}
\end{equation}
Minimizing Eq. \eqref{eq:m3} with respect to $n$ yields an equilibrium cluster size $n^\ast$
\begin{equation} \label {eq:m4}
\begin{aligned}
n^\ast = \frac {2a\lambda }{gE_r} 
\end{aligned}
\end{equation}
Approximating the line tension as $\lambda \approx E_a/\sigma$, the equilibrium cluster size is given as
\begin{equation} \label {eq:m5}
\begin{aligned}
n^\ast = \frac {\pi \sigma E_a}{2gE_r} 
\end{aligned}
\end{equation}

\begin{figure}
\centering
\includegraphics [scale=0.37]{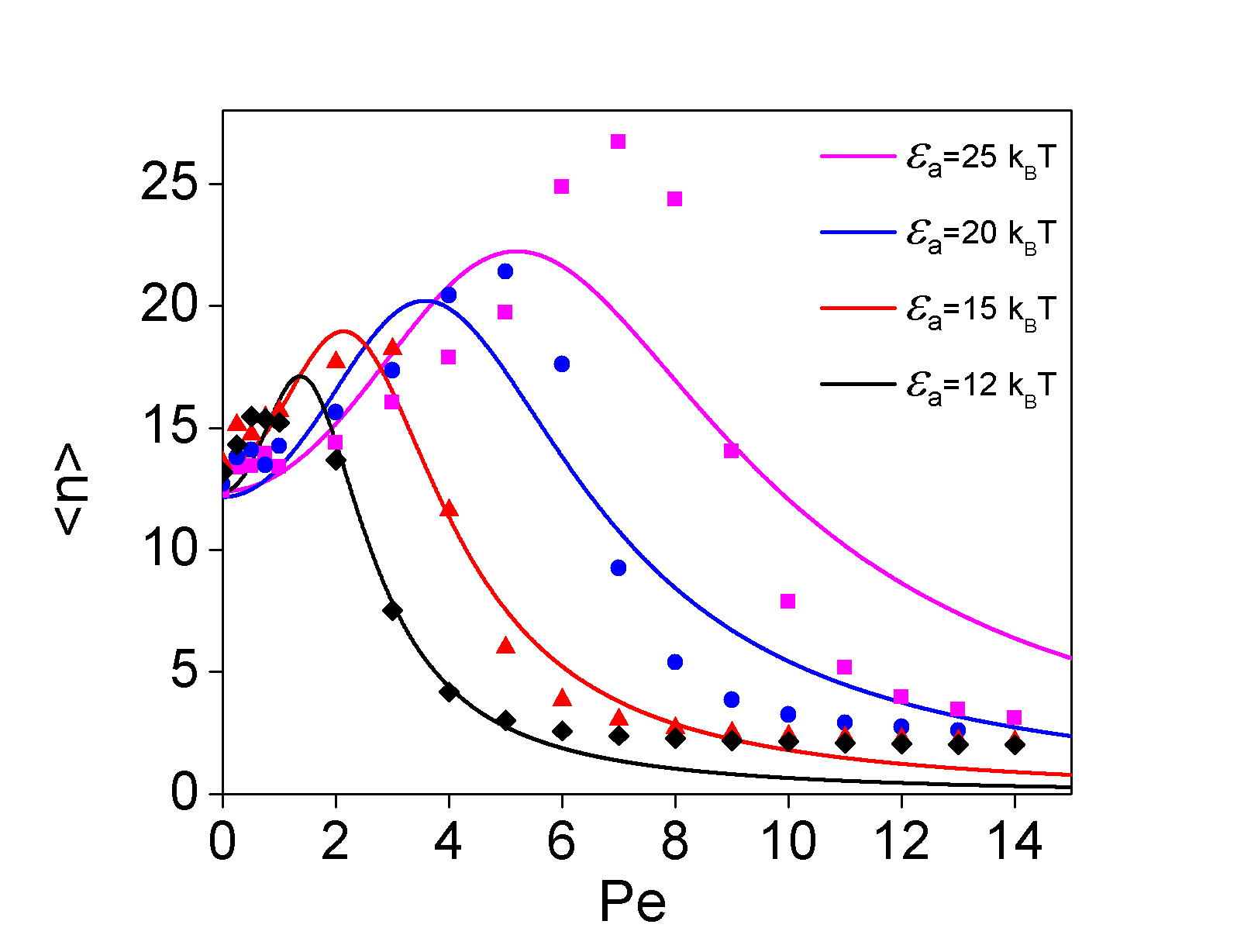}
\caption{(Colour online) Comparison between predictions of the model (Eq. \eqref{eq:m7}) with simulation data.}
 \label {fig:model}
\end{figure}

Eq. \eqref{eq:m5} gives the effect of interaction parameters on the equilibrium cluster size of passive particles with short-ranged attractive and long-ranged repulsive interactions. We use Eq. \eqref{eq:m5} as a starting point to analyze active particles, wherein each particle propels with a speed of $v$ and rotates freely in two dimensions in addition to their short-ranged attractive and long-ranged repulsive interparticle interactions. The critical issue is to know how to incorporate the effect of activity in terms of {\it effective\/} attractive and repulsive interactions in Eq. \eqref{eq:m5}. Consistent with previous work \cite{brader,poonpnas,berthier}, we propose that the role of activity in affecting inter-particle interactions has the following
features:
\begin{enumerate}
\item Both effective attraction and repulsion increase with $Pe$.
\item For small $Pe$, the increase of an activity-induced effective attraction is more pronounced than the activity-induced effective repulsion.
\item For large $Pe$, the effective activity-induced repulsion is getting more pronounced than the activity-induced effective attraction.
\end{enumerate}
Therefore, we replace $E_a$ in Eq. \eqref{eq:m5} to account for an effective activity-induced attraction as $E_a+aPe^p$ with two fit parameters: an amplitude $a$ and an exponent $p$. Similarly we add to $E_r$ an activity-induced repulsion, i.e. we replace $E_r$ by $E_r +bPe^q$ with two fit parameters, namely an
amplitude $b$ and an exponent $q$ which is larger than $p$. Therefore
\begin{equation} \label {eq:m7}
\begin{aligned}
n^\ast = \frac {\pi \sigma (E_a + aPe^p)}{2g(E_r+bPe^q)} 
\end{aligned}
\end{equation}
We find reasonable fits for our simulation data when fixing $p=2$ and $q=4$ adjusting only the amplitudes $a$ and $b$.
The actual fit parameters are given in Table I and in  Fig. \ref{fig:model}  the comparison between the model and simulation data is shown. Good fits are obtained for small $\epsilon_a$ but deviations are visible for larger  $\epsilon_a$. Here, the values of $E_a$ and $E_r$ correspond to the minimum and maximum of the potential described in Eq. \eqref{eq:1}. The parameter $g= 1.83$ is chosen such that the model predicts the size of the cluster size in the absence of activity. Hence this phenomenological model can give account for the trends  that at low $Pe$, $<n>$ increases and then decreases for high $Pe$ . 

\begin{table}[ht]
\caption{Fit parameters used in the effective equilibrium model}
\centering
\begin{tabular}{ccccc}
\hline
$\epsilon_a (k_BT)$ &\quad\quad $E_a (k_BT)$ &\quad\quad $E_r (k_BT)$ &\quad\quad $a$&\quad\quad $b$ \\ 
\hline
12 &\quad\quad 2.6 &\quad\quad 0.18 &\quad\quad 1.0605 &\quad\quad 0.0142  \\
15 &\quad\quad 3.3 &\quad\quad 0.21 &\quad\quad 0.5928 &\quad\quad 0.0030  \\
20 &\quad\quad 4.4 &\quad\quad 0.31 &\quad\quad 0.4550 &\quad\quad 0.0008  \\
25 &\quad\quad 5.5 &\quad\quad 0.38 &\quad\quad 0.3226 &\quad\quad 0.0002  \\
\hline
\end{tabular}
\end{table}

\section{Equilibrium versus dynamical clustering}
\begin{figure*}
\begin{center}
\begin{tabular}{ccc}
\includegraphics [scale=0.7]{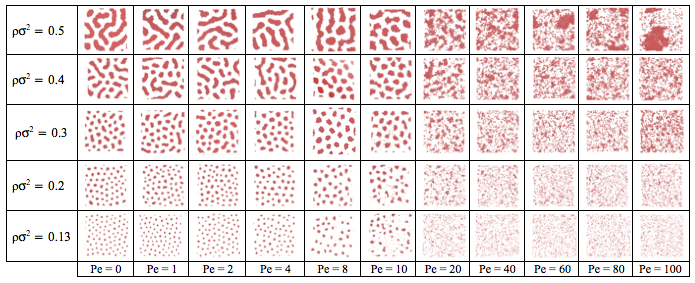}
\end{tabular}
\caption{(Colour online) Representative configurations of various states
of active particles with competing interactions for
varied Peclet number $Pe$ and reduced number density $\rho\sigma^2$ for $\epsilon_a=25 k_BT$ }
 \label {fig:phasediag}
 \end{center}
\end{figure*}
\begin{figure}
\begin{center}
\includegraphics [scale=0.75]{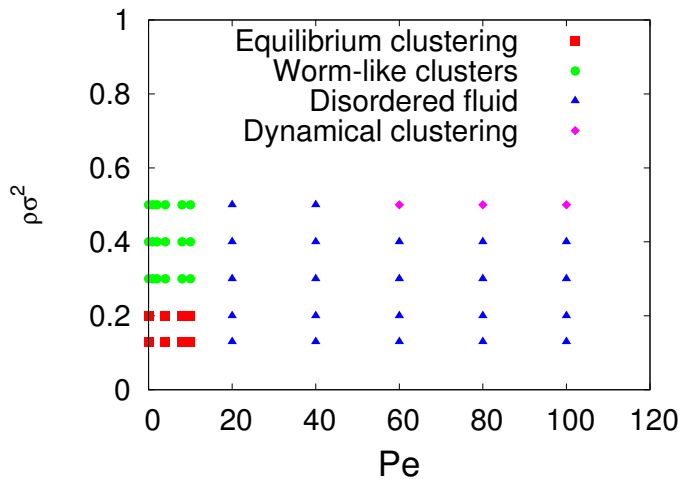}
\caption{(Colour online) State diagram of active particles with competing
interactions in the plane spanned by Peclet number $Pe$ and reduced
density $\rho\sigma^2$ for $\epsilon_a=25 k_BT$.}
 \label {fig:state}
 \end{center}
\end{figure}

Next we discuss the state behaviour of active particles with competing interactions to understand the link between equilibrium clustering and kinetic clustering induced by activity. Representative configurations obtained from simulations for different density and activity are shown in Fig. \ref{fig:phasediag} and a corresponding state diagram is presented in Fig. \ref{fig:state}. In the absence of activity ($Pe$=0), the particles form quasi-circular clusters at low density and extended worm-like clusters at higher densities. Upon increasing activity the phase behaviour changes significantly. For instance, up to a reduced density of 0.2, the activity increases the cluster sizes until a critical activity is reached and beyond this activity, clusters dissolve leading to a disordered fluid phase. In the density range of 0.3 to 0.4, we see a similar behaviour but now the elongated worm-like clusters are getting smaller in length upon increasing the activity. Further increase in activity again leads to a disordered fluid in this density regime. At higher densities such as 0.5, the disordered fluid is followed by phase-separation due to kinetic clustering induced by motility as found in earlier reports on purely repulsive particle \cite{buttinoni,bialkeepl2013}. 
At these high Peclet numbers, details of the interparticle interaction except for the repulsive core are not
relevant. The dynamic clustering observed at high $Pe$ is therefore explained as a pure kinetic effect.
It is well separated from the equilibrium clustering considered earlier which occurs at small $Pe$ 
and small densities demonstrating that these two clustering effects are qualitatively distinct.   

\begin{figure}[t]
\begin{center}
\includegraphics [scale=0.35]{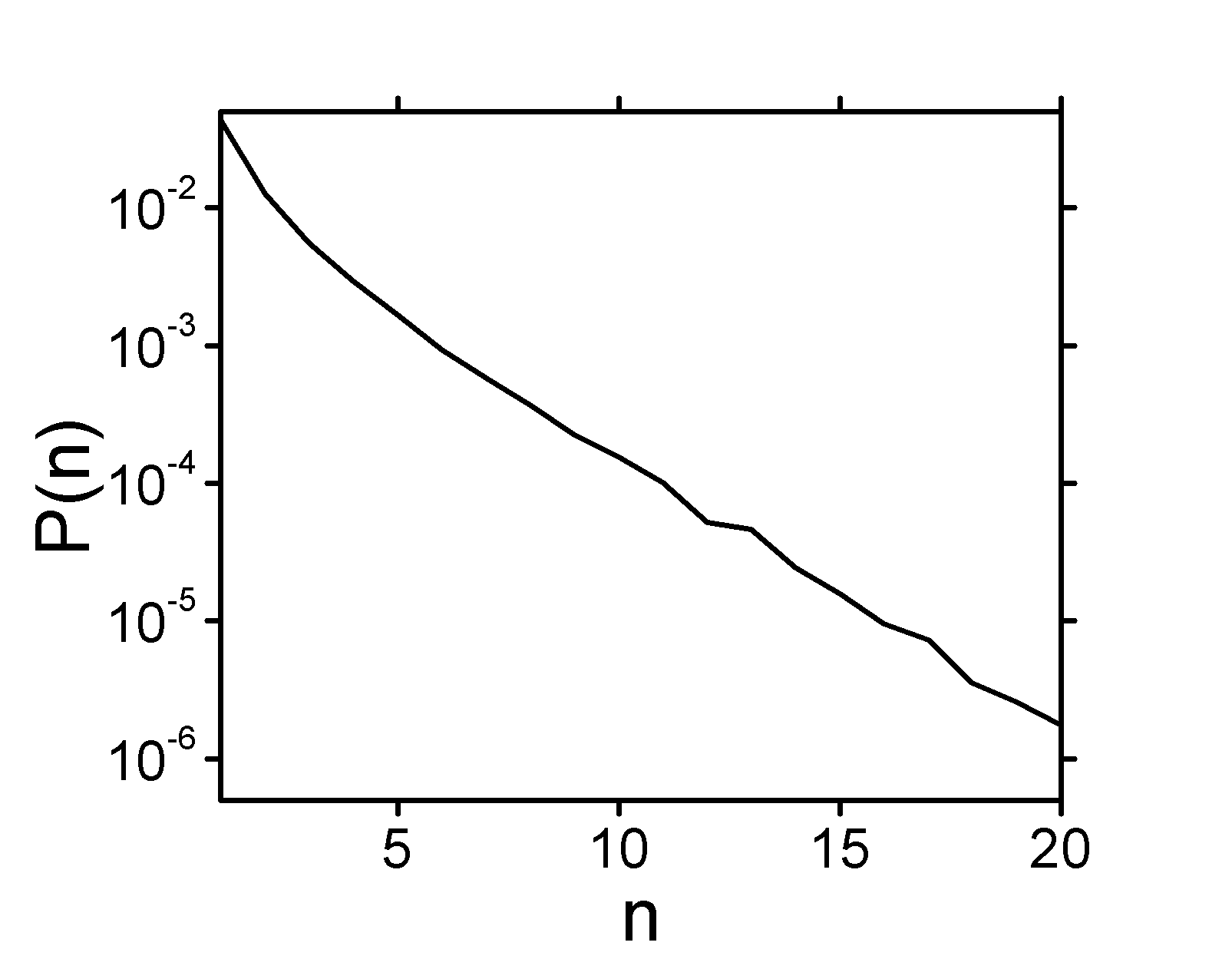}
\caption{Cluster size distribution for $Pe$ = 100, density $\rho\sigma^2$ = 0.13 and ${\epsilon_a} = 25k_BT$.} 
 \label {fig:pofn100}
 \end{center}
\end{figure}

We now revisit the cluster size distribution function $P(n)$ in the limit of high Peclet numbers
where only the repulsive core is relevant, which is then to be compared to the previous data shown in Fig. \ref{fig:pe0}. Results for $Pe=100$ are shown in Fig. \ref{fig:pofn100} on a semi-log scale. The average cluster size in this case is 1.8. The clusters are distinctly different, both in size and shape, from that observed in equilibrium (Fig. \ref{fig:pe0}a) and at moderate activity (Fig. \ref{fig:pe0}b), although $\epsilon_a$ and density are the same. This means that activity can be used as a knob, to an extent, in tuning the cluster size either to increase or to decrease depending upon the fixed parameters of interactions. This unique feature is not present for purely motility-induced clustering in repulsive colloids. In more detail, as can be deduced from Fig. \ref{fig:pofn100}, the cluster size distribution $P(n)$ is almost linear in the semi-logarithmic plot, except for small $n$ ($n < 4$). This shows that there is an exponential decay in $P(n)$ for large $n$. The data for $n < 5$ are compatible with a power-law scaling. Majumdar et al. \cite{majumdar} showed that for an effective aggregating and fragmenting particle system $P(n)$ decays exponentially for low aggregation rates and decays as power-law for high aggregation rates. In the present case, the initial power-law decay denoting an effective aggregation process indicates motility-induced kinetic clustering at high Peclet numbers. The exponential decay denoting effective fragmentation indicates fragmentation due to activity. Therefore, high activity induces both aggregating and fragmenting processes. In contrast, in the case of $Pe$ = 6, where both the interaction potential and activity affect clustering, the cluster size distribution $P(n)$ 
does not show any conclusive scaling. 

Furthermore we note that our results are different from flying crystals which were found to exist with and without cohesive forces \cite{chate,toner,menzel}. In these studies, aligning forces between the particles are relevant.  In our simulations, we have not considered aligning forces and therefore we have not observed flying crystals. The velocity vectors of the particles in the clusters obtained from our simulations are randomly oriented as the particles are freely rotating with their rotational diffusion coefficient. Therefore, there is only a random  and undirected migration of the clusters. 
\section{Conclusion}
In conclusion, by using Brownian dynamics computer simulations,
 we explored how equilibrium clustering  is affected for self-propelled
colloidal particles. While this clustering process is stable under self-propulsion, depending on the values of interaction parameter, the cluster size can initially increase with the strength of self-propulsion,
before it decreases for large activity. This allows to controll the strength of
active clustering via the interparticle interactions. A phenomenological model is shown to qualitatively explain the non-monotonic variation of cluster size with activity. For the future, it would be interesting to construct a dynamical density functional theory
for the clustering considered here by unifying the density functional theory
designed for equilibrium clustering \cite {wu} with
that designed for kinetic clustering \cite{bialkeepl2013}. Moreover,
hydrodynamic effects should be explored by more sophisticated simulation models \cite{stark}.
Furthermore our predictions can in principle be verified in experiments on synthetic or bacterial microswimmers
with well-defined interactions. In particular the combination of depletants, particle charge and magnetic
dipole moments \cite{baraban1,baraban2}
opens new ways to steer the interparticle interactions between swimmers and therefore the
details of the clustering behaviour.
\acknowledgments
We thank Julian Bialk\'e, Andreas Kaiser and Borge ten Hagen for helpful discussions.
This work was supported by the DAAD and by the ERC Advanced Grand INTERCOCOS (Grand No. 267499).
\bibliographystyle{aipnum4-1}
\bibliography{references}

\begin{thebibliography}{42}%
\makeatletter
\providecommand \@ifxundefined [1]{%
 \@ifx{#1\undefined}
}%
\providecommand \@ifnum [1]{%
 \ifnum #1\expandafter \@firstoftwo
 \else \expandafter \@secondoftwo
 \fi
}%
\providecommand \@ifx [1]{%
 \ifx #1\expandafter \@firstoftwo
 \else \expandafter \@secondoftwo
 \fi
}%
\providecommand \natexlab [1]{#1}%
\providecommand \enquote  [1]{``#1''}%
\providecommand \bibnamefont  [1]{#1}%
\providecommand \bibfnamefont [1]{#1}%
\providecommand \citenamefont [1]{#1}%
\providecommand \href@noop [0]{\@secondoftwo}%
\providecommand \href [0]{\begingroup \@sanitize@url \@href}%
\providecommand \@href[1]{\@@startlink{#1}\@@href}%
\providecommand \@@href[1]{\endgroup#1\@@endlink}%
\providecommand \@sanitize@url [0]{\catcode `\\12\catcode `\$12\catcode
  `\&12\catcode `\#12\catcode `\^12\catcode `\_12\catcode `\%12\relax}%
\providecommand \@@startlink[1]{}%
\providecommand \@@endlink[0]{}%
\providecommand \url  [0]{\begingroup\@sanitize@url \@url }%
\providecommand \@url [1]{\endgroup\@href {#1}{\urlprefix }}%
\providecommand \urlprefix  [0]{URL }%
\providecommand \Eprint [0]{\href }%
\providecommand \doibase [0]{http://dx.doi.org/}%
\providecommand \selectlanguage [0]{\@gobble}%
\providecommand \bibinfo  [0]{\@secondoftwo}%
\providecommand \bibfield  [0]{\@secondoftwo}%
\providecommand \translation [1]{[#1]}%
\providecommand \BibitemOpen [0]{}%
\providecommand \bibitemStop [0]{}%
\providecommand \bibitemNoStop [0]{.\EOS\space}%
\providecommand \EOS [0]{\spacefactor3000\relax}%
\providecommand \BibitemShut  [1]{\csname bibitem#1\endcsname}%
\let\auto@bib@innerbib\@empty
\bibitem [{\citenamefont {Groenewold}\ and\ \citenamefont
  {Kegel}(2001)}]{kegel}%
  \BibitemOpen
  \bibfield  {author} {\bibinfo {author} {\bibfnamefont {J.}~\bibnamefont
  {Groenewold}}\ and\ \bibinfo {author} {\bibfnamefont {W.~K.}\ \bibnamefont
  {Kegel}},\ }\href@noop {} {\bibfield  {journal} {\bibinfo  {journal} {J.
  Phys. Chem. B}\ }\textbf {\bibinfo {volume} {105}},\ \bibinfo {pages} {11702}
  (\bibinfo {year} {2001})}\BibitemShut {NoStop}%
\bibitem [{\citenamefont {Sciortino}\ \emph {et~al.}(2004)\citenamefont
  {Sciortino}, \citenamefont {Mossa}, \citenamefont {Zaccarelli},\ and\
  \citenamefont {Tartaglia}}]{sciortino}%
  \BibitemOpen
  \bibfield  {author} {\bibinfo {author} {\bibfnamefont {F.}~\bibnamefont
  {Sciortino}}, \bibinfo {author} {\bibfnamefont {S.}~\bibnamefont {Mossa}},
  \bibinfo {author} {\bibfnamefont {E.}~\bibnamefont {Zaccarelli}}, \ and\
  \bibinfo {author} {\bibfnamefont {P.}~\bibnamefont {Tartaglia}},\ }\href
  {\doibase 10.1103/PhysRevLett.93.055701} {\bibfield  {journal} {\bibinfo
  {journal} {Phys. Rev. Lett.}\ }\textbf {\bibinfo {volume} {93}},\ \bibinfo
  {pages} {055701} (\bibinfo {year} {2004})}\BibitemShut {NoStop}%
\bibitem [{\citenamefont {Imperio}\ and\ \citenamefont
  {Reatto}(2006)}]{imperio}%
  \BibitemOpen
  \bibfield  {author} {\bibinfo {author} {\bibfnamefont {A.}~\bibnamefont
  {Imperio}}\ and\ \bibinfo {author} {\bibfnamefont {L.}~\bibnamefont
  {Reatto}},\ }\href {\doibase http://dx.doi.org/10.1063/1.2185618} {\bibfield
  {journal} {\bibinfo  {journal} {J. Chem. Phys.}\ }\textbf {\bibinfo {volume}
  {124}},\ \bibinfo {eid} {164712} (\bibinfo {year} {2006})}\BibitemShut
  {NoStop}%
\bibitem [{\citenamefont {Godfrin}\ \emph {et~al.}(2013)\citenamefont
  {Godfrin}, \citenamefont {Castañeda-Priego}, \citenamefont {Liu},\ and\
  \citenamefont {Wagner}}]{godfrin}%
  \BibitemOpen
  \bibfield  {author} {\bibinfo {author} {\bibfnamefont {P.~D.}\ \bibnamefont
  {Godfrin}}, \bibinfo {author} {\bibfnamefont {R.}~\bibnamefont
  {Castañeda-Priego}}, \bibinfo {author} {\bibfnamefont {Y.}~\bibnamefont
  {Liu}}, \ and\ \bibinfo {author} {\bibfnamefont {N.~J.}\ \bibnamefont
  {Wagner}},\ }\href@noop {} {\bibfield  {journal} {\bibinfo  {journal} {J.
  Chem. Phys.}\ }\textbf {\bibinfo {volume} {139}},\ \bibinfo {eid} {154904}
  (\bibinfo {year} {2013})}\BibitemShut {NoStop}%
\bibitem [{\citenamefont {Mani}\ \emph {et~al.}(2014)\citenamefont {Mani},
  \citenamefont {Lechner}, \citenamefont {Kegel},\ and\ \citenamefont
  {Bolhuis}}]{mani}%
  \BibitemOpen
  \bibfield  {author} {\bibinfo {author} {\bibfnamefont {E.}~\bibnamefont
  {Mani}}, \bibinfo {author} {\bibfnamefont {W.}~\bibnamefont {Lechner}},
  \bibinfo {author} {\bibfnamefont {W.~K.}\ \bibnamefont {Kegel}}, \ and\
  \bibinfo {author} {\bibfnamefont {P.~G.}\ \bibnamefont {Bolhuis}},\ }\href
  {\doibase 10.1039/C3SM53058B} {\bibfield  {journal} {\bibinfo  {journal}
  {Soft Matter}\ }\textbf {\bibinfo {volume} {10}},\ \bibinfo {pages} {4479}
  (\bibinfo {year} {2014})}\BibitemShut {NoStop}%
\bibitem [{\citenamefont {Stradner}\ \emph {et~al.}(2004)\citenamefont
  {Stradner}, \citenamefont {Sedgwick}, \citenamefont {Cardinaux},
  \citenamefont {Poon}, \citenamefont {Egelhaaf},\ and\ \citenamefont
  {Schurtenberger}}]{stradner}%
  \BibitemOpen
  \bibfield  {author} {\bibinfo {author} {\bibfnamefont {A.}~\bibnamefont
  {Stradner}}, \bibinfo {author} {\bibfnamefont {H.}~\bibnamefont {Sedgwick}},
  \bibinfo {author} {\bibfnamefont {F.}~\bibnamefont {Cardinaux}}, \bibinfo
  {author} {\bibfnamefont {W.~C.~K.}\ \bibnamefont {Poon}}, \bibinfo {author}
  {\bibfnamefont {S.~U.}\ \bibnamefont {Egelhaaf}}, \ and\ \bibinfo {author}
  {\bibfnamefont {P.}~\bibnamefont {Schurtenberger}},\ }\href
  {http://dx.doi.org/10.1038/nature03109} {\bibfield  {journal} {\bibinfo
  {journal} {Nature}\ }\textbf {\bibinfo {volume} {432}},\ \bibinfo {pages}
  {492} (\bibinfo {year} {2004})}\BibitemShut {NoStop}%
\bibitem [{\citenamefont {Campbell}\ \emph {et~al.}(2005)\citenamefont
  {Campbell}, \citenamefont {Anderson}, \citenamefont {van Duijneveldt},\ and\
  \citenamefont {Bartlett}}]{bartlett}%
  \BibitemOpen
  \bibfield  {author} {\bibinfo {author} {\bibfnamefont {A.~I.}\ \bibnamefont
  {Campbell}}, \bibinfo {author} {\bibfnamefont {V.~J.}\ \bibnamefont
  {Anderson}}, \bibinfo {author} {\bibfnamefont {J.~S.}\ \bibnamefont {van
  Duijneveldt}}, \ and\ \bibinfo {author} {\bibfnamefont {P.}~\bibnamefont
  {Bartlett}},\ }\href {\doibase 10.1103/PhysRevLett.94.208301} {\bibfield
  {journal} {\bibinfo  {journal} {Phys. Rev. Lett.}\ }\textbf {\bibinfo
  {volume} {94}},\ \bibinfo {pages} {208301} (\bibinfo {year}
  {2005})}\BibitemShut {NoStop}%
\bibitem [{\citenamefont {Romanczuk}\ \emph {et~al.}(2008)\citenamefont
  {Romanczuk}, \citenamefont {Erdmann}, \citenamefont {Engel},\ and\
  \citenamefont {Schimansky-Geier}}]{schimansky}%
  \BibitemOpen
  \bibfield  {author} {\bibinfo {author} {\bibfnamefont {P.}~\bibnamefont
  {Romanczuk}}, \bibinfo {author} {\bibfnamefont {U.}~\bibnamefont {Erdmann}},
  \bibinfo {author} {\bibfnamefont {H.}~\bibnamefont {Engel}}, \ and\ \bibinfo
  {author} {\bibfnamefont {L.}~\bibnamefont {Schimansky-Geier}},\ }\href@noop
  {} {\bibfield  {journal} {\bibinfo  {journal} {Eur. Phys. J.: Spec. Top.}\
  }\textbf {\bibinfo {volume} {157}},\ \bibinfo {pages} {61} (\bibinfo {year}
  {2008})}\BibitemShut {NoStop}%
\bibitem [{\citenamefont {Marchetti}\ \emph {et~al.}(2013)\citenamefont
  {Marchetti}, \citenamefont {Joanny}, \citenamefont {Ramaswamy}, \citenamefont
  {Liverpool}, \citenamefont {Prost}, \citenamefont {Rao},\ and\ \citenamefont
  {Simha}}]{marchetti}%
  \BibitemOpen
  \bibfield  {author} {\bibinfo {author} {\bibfnamefont {M.}~\bibnamefont
  {Marchetti}}, \bibinfo {author} {\bibfnamefont {J.}~\bibnamefont {Joanny}},
  \bibinfo {author} {\bibfnamefont {S.}~\bibnamefont {Ramaswamy}}, \bibinfo
  {author} {\bibfnamefont {T.}~\bibnamefont {Liverpool}}, \bibinfo {author}
  {\bibfnamefont {J.}~\bibnamefont {Prost}}, \bibinfo {author} {\bibfnamefont
  {M.}~\bibnamefont {Rao}}, \ and\ \bibinfo {author} {\bibfnamefont
  {R.}~\bibnamefont {Simha}},\ }\href@noop {} {\bibfield  {journal} {\bibinfo
  {journal} {Rev. Mod. Phys.}\ }\textbf {\bibinfo {volume} {85}},\ \bibinfo
  {pages} {1143} (\bibinfo {year} {2013})}\BibitemShut {NoStop}%
\bibitem [{\citenamefont {Bialk\'e}, \citenamefont {Speck},\ and\ \citenamefont
  {L\"owen}(2015)}]{review}%
  \BibitemOpen
  \bibfield  {author} {\bibinfo {author} {\bibfnamefont {J.}~\bibnamefont
  {Bialk\'e}}, \bibinfo {author} {\bibfnamefont {T.}~\bibnamefont {Speck}}, \
  and\ \bibinfo {author} {\bibfnamefont {H.}~\bibnamefont {L\"owen}},\ }\href
  {\doibase http://dx.doi.org/10.1016/j.jnoncrysol.2014.08.011} {\bibfield
  {journal} {\bibinfo  {journal} {Journal of Non-Crystalline Solids}\ }\textbf
  {\bibinfo {volume} {407}},\ \bibinfo {pages} {367 } (\bibinfo {year}
  {2015})}\BibitemShut {NoStop}%
\bibitem [{\citenamefont {Ginot}\ \emph {et~al.}(2015)\citenamefont {Ginot},
  \citenamefont {Theurkauff}, \citenamefont {Levis}, \citenamefont {Ybert},
  \citenamefont {Bocquet}, \citenamefont {Berthier},\ and\ \citenamefont
  {Cottin-Bizonne}}]{berthier}%
  \BibitemOpen
  \bibfield  {author} {\bibinfo {author} {\bibfnamefont {F.}~\bibnamefont
  {Ginot}}, \bibinfo {author} {\bibfnamefont {I.}~\bibnamefont {Theurkauff}},
  \bibinfo {author} {\bibfnamefont {D.}~\bibnamefont {Levis}}, \bibinfo
  {author} {\bibfnamefont {C.}~\bibnamefont {Ybert}}, \bibinfo {author}
  {\bibfnamefont {L.}~\bibnamefont {Bocquet}}, \bibinfo {author} {\bibfnamefont
  {L.}~\bibnamefont {Berthier}}, \ and\ \bibinfo {author} {\bibfnamefont
  {C.}~\bibnamefont {Cottin-Bizonne}},\ }\href {\doibase
  10.1103/PhysRevX.5.011004} {\bibfield  {journal} {\bibinfo  {journal} {Phys.
  Rev. X}\ }\textbf {\bibinfo {volume} {5}},\ \bibinfo {pages} {011004}
  (\bibinfo {year} {2015})}\BibitemShut {NoStop}%
\bibitem [{\citenamefont {Buttinoni}\ \emph {et~al.}(2013)\citenamefont
  {Buttinoni}, \citenamefont {Bialk\'e}, \citenamefont {K\"ummel},
  \citenamefont {L\"owen}, \citenamefont {Bechinger},\ and\ \citenamefont
  {Speck}}]{buttinoni}%
  \BibitemOpen
  \bibfield  {author} {\bibinfo {author} {\bibfnamefont {I.}~\bibnamefont
  {Buttinoni}}, \bibinfo {author} {\bibfnamefont {J.}~\bibnamefont {Bialk\'e}},
  \bibinfo {author} {\bibfnamefont {F.}~\bibnamefont {K\"ummel}}, \bibinfo
  {author} {\bibfnamefont {H.}~\bibnamefont {L\"owen}}, \bibinfo {author}
  {\bibfnamefont {C.}~\bibnamefont {Bechinger}}, \ and\ \bibinfo {author}
  {\bibfnamefont {T.}~\bibnamefont {Speck}},\ }\href
  {http://link.aps.org/doi/10.1103/PhysRevLett.110.238301} {\bibfield
  {journal} {\bibinfo  {journal} {Phys. Rev. Lett.}\ }\textbf {\bibinfo
  {volume} {110}},\ \bibinfo {pages} {238301} (\bibinfo {year}
  {2013})}\BibitemShut {NoStop}%
\bibitem [{\citenamefont {Palacci}\ \emph {et~al.}(2013)\citenamefont
  {Palacci}, \citenamefont {Sacanna}, \citenamefont {Steinberg}, \citenamefont
  {Pine},\ and\ \citenamefont {Chaikin}}]{chaikin}%
  \BibitemOpen
  \bibfield  {author} {\bibinfo {author} {\bibfnamefont {J.}~\bibnamefont
  {Palacci}}, \bibinfo {author} {\bibfnamefont {S.}~\bibnamefont {Sacanna}},
  \bibinfo {author} {\bibfnamefont {A.}~\bibnamefont {Steinberg}}, \bibinfo
  {author} {\bibfnamefont {D.}~\bibnamefont {Pine}}, \ and\ \bibinfo {author}
  {\bibfnamefont {P.}~\bibnamefont {Chaikin}},\ }\href@noop {} {\bibfield
  {journal} {\bibinfo  {journal} {Science}\ }\textbf {\bibinfo {volume}
  {339}},\ \bibinfo {pages} {936} (\bibinfo {year} {2013})}\BibitemShut
  {NoStop}%
\bibitem [{\citenamefont {Fily}\ and\ \citenamefont
  {Marchetti}(2012)}]{fily2012}%
  \BibitemOpen
  \bibfield  {author} {\bibinfo {author} {\bibfnamefont {Y.}~\bibnamefont
  {Fily}}\ and\ \bibinfo {author} {\bibfnamefont {M.~C.}\ \bibnamefont
  {Marchetti}},\ }\href {\doibase 10.1103/PhysRevLett.108.235702} {\bibfield
  {journal} {\bibinfo  {journal} {Phys. Rev. Lett.}\ }\textbf {\bibinfo
  {volume} {108}},\ \bibinfo {pages} {235702} (\bibinfo {year}
  {2012})}\BibitemShut {NoStop}%
\bibitem [{\citenamefont {Stenhammar}\ \emph {et~al.}(2013)\citenamefont
  {Stenhammar}, \citenamefont {Tiribocchi}, \citenamefont {Allen},
  \citenamefont {Marenduzzo},\ and\ \citenamefont {Cates}}]{stenhammar2013}%
  \BibitemOpen
  \bibfield  {author} {\bibinfo {author} {\bibfnamefont {J.}~\bibnamefont
  {Stenhammar}}, \bibinfo {author} {\bibfnamefont {A.}~\bibnamefont
  {Tiribocchi}}, \bibinfo {author} {\bibfnamefont {R.~J.}\ \bibnamefont
  {Allen}}, \bibinfo {author} {\bibfnamefont {D.}~\bibnamefont {Marenduzzo}}, \
  and\ \bibinfo {author} {\bibfnamefont {M.~E.}\ \bibnamefont {Cates}},\ }\href
  {\doibase 10.1103/PhysRevLett.111.145702} {\bibfield  {journal} {\bibinfo
  {journal} {Phys. Rev. Lett.}\ }\textbf {\bibinfo {volume} {111}},\ \bibinfo
  {pages} {145702} (\bibinfo {year} {2013})}\BibitemShut {NoStop}%
\bibitem [{\citenamefont {Redner}, \citenamefont {Hagan},\ and\ \citenamefont
  {Baskaran}(2013)}]{redner}%
  \BibitemOpen
  \bibfield  {author} {\bibinfo {author} {\bibfnamefont {G.~S.}\ \bibnamefont
  {Redner}}, \bibinfo {author} {\bibfnamefont {M.~F.}\ \bibnamefont {Hagan}}, \
  and\ \bibinfo {author} {\bibfnamefont {A.}~\bibnamefont {Baskaran}},\ }\href
  {\doibase 10.1103/PhysRevLett.110.055701} {\bibfield  {journal} {\bibinfo
  {journal} {Phys. Rev. Lett.}\ }\textbf {\bibinfo {volume} {110}},\ \bibinfo
  {pages} {055701} (\bibinfo {year} {2013})}\BibitemShut {NoStop}%
\bibitem [{\citenamefont {Bialk{\'e}}, \citenamefont {L{\"o}wen},\ and\
  \citenamefont {Speck}(2013)}]{bialkeepl2013}%
  \BibitemOpen
  \bibfield  {author} {\bibinfo {author} {\bibfnamefont {J.}~\bibnamefont
  {Bialk{\'e}}}, \bibinfo {author} {\bibfnamefont {H.}~\bibnamefont
  {L{\"o}wen}}, \ and\ \bibinfo {author} {\bibfnamefont {T.}~\bibnamefont
  {Speck}},\ }\href {http://stacks.iop.org/0295-5075/103/i=3/a=30008}
  {\bibfield  {journal} {\bibinfo  {journal} {Eur. Phys. Lett.}\ }\textbf
  {\bibinfo {volume} {103}},\ \bibinfo {pages} {30008} (\bibinfo {year}
  {2013})}\BibitemShut {NoStop}%
\bibitem [{\citenamefont {Stenhammar}\ \emph {et~al.}(2014)\citenamefont
  {Stenhammar}, \citenamefont {Marenduzzo}, \citenamefont {Allen},\ and\
  \citenamefont {Cates}}]{stenhammar2014}%
  \BibitemOpen
  \bibfield  {author} {\bibinfo {author} {\bibfnamefont {J.}~\bibnamefont
  {Stenhammar}}, \bibinfo {author} {\bibfnamefont {D.}~\bibnamefont
  {Marenduzzo}}, \bibinfo {author} {\bibfnamefont {R.~J.}\ \bibnamefont
  {Allen}}, \ and\ \bibinfo {author} {\bibfnamefont {M.~E.}\ \bibnamefont
  {Cates}},\ }\href {\doibase 10.1039/C3SM52813H} {\bibfield  {journal}
  {\bibinfo  {journal} {Soft Matter}\ }\textbf {\bibinfo {volume} {10}},\
  \bibinfo {pages} {1489} (\bibinfo {year} {2014})}\BibitemShut {NoStop}%
\bibitem [{\citenamefont {Fily}, \citenamefont {Henkes},\ and\ \citenamefont
  {Marchetti}(2014)}]{fily2014}%
  \BibitemOpen
  \bibfield  {author} {\bibinfo {author} {\bibfnamefont {Y.}~\bibnamefont
  {Fily}}, \bibinfo {author} {\bibfnamefont {S.}~\bibnamefont {Henkes}}, \ and\
  \bibinfo {author} {\bibfnamefont {M.~C.}\ \bibnamefont {Marchetti}},\ }\href
  {\doibase 10.1039/C3SM52469H} {\bibfield  {journal} {\bibinfo  {journal}
  {Soft Matter}\ }\textbf {\bibinfo {volume} {10}},\ \bibinfo {pages} {2132}
  (\bibinfo {year} {2014})}\BibitemShut {NoStop}%
\bibitem [{\citenamefont {Speck}\ \emph {et~al.}(2014)\citenamefont {Speck},
  \citenamefont {Bialk\'e}, \citenamefont {Menzel},\ and\ \citenamefont
  {L\"owen}}]{speck}%
  \BibitemOpen
  \bibfield  {author} {\bibinfo {author} {\bibfnamefont {T.}~\bibnamefont
  {Speck}}, \bibinfo {author} {\bibfnamefont {J.}~\bibnamefont {Bialk\'e}},
  \bibinfo {author} {\bibfnamefont {A.~M.}\ \bibnamefont {Menzel}}, \ and\
  \bibinfo {author} {\bibfnamefont {H.}~\bibnamefont {L\"owen}},\ }\href
  {\doibase 10.1103/PhysRevLett.112.218304} {\bibfield  {journal} {\bibinfo
  {journal} {Phys. Rev. Lett.}\ }\textbf {\bibinfo {volume} {112}},\ \bibinfo
  {pages} {218304} (\bibinfo {year} {2014})}\BibitemShut {NoStop}%
\bibitem [{\citenamefont {Wysocki}, \citenamefont {Winkler},\ and\
  \citenamefont {Gompper}(2014)}]{wysocki}%
  \BibitemOpen
  \bibfield  {author} {\bibinfo {author} {\bibfnamefont {A.}~\bibnamefont
  {Wysocki}}, \bibinfo {author} {\bibfnamefont {R.~G.}\ \bibnamefont
  {Winkler}}, \ and\ \bibinfo {author} {\bibfnamefont {G.}~\bibnamefont
  {Gompper}},\ }\href {http://stacks.iop.org/0295-5075/105/i=4/a=48004}
  {\bibfield  {journal} {\bibinfo  {journal} {Eur. Phys. Lett.}\ }\textbf
  {\bibinfo {volume} {105}},\ \bibinfo {pages} {48004} (\bibinfo {year}
  {2014})}\BibitemShut {NoStop}%
\bibitem [{\citenamefont {Mognetti}\ \emph {et~al.}(2013)\citenamefont
  {Mognetti}, \citenamefont {\ifmmode \check{S}\else
  \v{S}\fi{}ari\ifmmode~\acute{c}\else \'{c}\fi{}}, \citenamefont
  {Angioletti-Uberti}, \citenamefont {Cacciuto}, \citenamefont {Valeriani},\
  and\ \citenamefont {Frenkel}}]{mognetti}%
  \BibitemOpen
  \bibfield  {author} {\bibinfo {author} {\bibfnamefont {B.~M.}\ \bibnamefont
  {Mognetti}}, \bibinfo {author} {\bibfnamefont {A.}~\bibnamefont {\ifmmode
  \check{S}\else \v{S}\fi{}ari\ifmmode~\acute{c}\else \'{c}\fi{}}}, \bibinfo
  {author} {\bibfnamefont {S.}~\bibnamefont {Angioletti-Uberti}}, \bibinfo
  {author} {\bibfnamefont {A.}~\bibnamefont {Cacciuto}}, \bibinfo {author}
  {\bibfnamefont {C.}~\bibnamefont {Valeriani}}, \ and\ \bibinfo {author}
  {\bibfnamefont {D.}~\bibnamefont {Frenkel}},\ }\href {\doibase
  10.1103/PhysRevLett.111.245702} {\bibfield  {journal} {\bibinfo  {journal}
  {Phys. Rev. Lett.}\ }\textbf {\bibinfo {volume} {111}},\ \bibinfo {pages}
  {245702} (\bibinfo {year} {2013})}\BibitemShut {NoStop}%
\bibitem [{\citenamefont {Tailleur}\ and\ \citenamefont
  {Cates}(2008)}]{tailleur}%
  \BibitemOpen
  \bibfield  {author} {\bibinfo {author} {\bibfnamefont {J.}~\bibnamefont
  {Tailleur}}\ and\ \bibinfo {author} {\bibfnamefont {M.~E.}\ \bibnamefont
  {Cates}},\ }\href {\doibase 10.1103/PhysRevLett.100.218103} {\bibfield
  {journal} {\bibinfo  {journal} {Phys. Rev. Lett.}\ }\textbf {\bibinfo
  {volume} {100}},\ \bibinfo {pages} {218103} (\bibinfo {year}
  {2008})}\BibitemShut {NoStop}%
\bibitem [{\citenamefont {Cates}\ \emph {et~al.}(2010)\citenamefont {Cates},
  \citenamefont {Marenduzzo}, \citenamefont {Pagonabarraga},\ and\
  \citenamefont {Tailleur}}]{cates2010}%
  \BibitemOpen
  \bibfield  {author} {\bibinfo {author} {\bibfnamefont {M.~E.}\ \bibnamefont
  {Cates}}, \bibinfo {author} {\bibfnamefont {D.}~\bibnamefont {Marenduzzo}},
  \bibinfo {author} {\bibfnamefont {I.}~\bibnamefont {Pagonabarraga}}, \ and\
  \bibinfo {author} {\bibfnamefont {J.}~\bibnamefont {Tailleur}},\ }\href
  {\doibase 10.1073/pnas.1001994107} {\bibfield  {journal} {\bibinfo  {journal}
  {Proc. Natl. Acad. Sci. USA}\ }\textbf {\bibinfo {volume} {107}},\ \bibinfo
  {pages} {11715} (\bibinfo {year} {2010})}\BibitemShut {NoStop}%
\bibitem [{\citenamefont {Cates}\ and\ \citenamefont
  {Tailleur}(2013)}]{cates2013}%
  \BibitemOpen
  \bibfield  {author} {\bibinfo {author} {\bibfnamefont {M.~E.}\ \bibnamefont
  {Cates}}\ and\ \bibinfo {author} {\bibfnamefont {J.}~\bibnamefont
  {Tailleur}},\ }\href {http://stacks.iop.org/0295-5075/101/i=2/a=20010}
  {\bibfield  {journal} {\bibinfo  {journal} {Eur. Phys. Lett.}\ }\textbf
  {\bibinfo {volume} {101}},\ \bibinfo {pages} {20010} (\bibinfo {year}
  {2013})}\BibitemShut {NoStop}%
\bibitem [{\citenamefont {Wittkowski}\ \emph {et~al.}(2014)\citenamefont
  {Wittkowski}, \citenamefont {Tiribocchi}, \citenamefont {Stenhammar},
  \citenamefont {Allen}, \citenamefont {Marenduzzo},\ and\ \citenamefont
  {Cates}}]{wittkowski}%
  \BibitemOpen
  \bibfield  {author} {\bibinfo {author} {\bibfnamefont {R.}~\bibnamefont
  {Wittkowski}}, \bibinfo {author} {\bibfnamefont {A.}~\bibnamefont
  {Tiribocchi}}, \bibinfo {author} {\bibfnamefont {J.}~\bibnamefont
  {Stenhammar}}, \bibinfo {author} {\bibfnamefont {R.~J.}\ \bibnamefont
  {Allen}}, \bibinfo {author} {\bibfnamefont {D.}~\bibnamefont {Marenduzzo}}, \
  and\ \bibinfo {author} {\bibfnamefont {M.~E.}\ \bibnamefont {Cates}},\ }\href
  {http://dx.doi.org/10.1038/ncomms5351} {\bibfield  {journal} {\bibinfo
  {journal} {Nature Commun}\ }\textbf {\bibinfo {volume} {5}} (\bibinfo {year}
  {2014})}\BibitemShut {NoStop}%
\bibitem [{\citenamefont {Peruani}\ and\ \citenamefont
  {B{\"a}r}(2013)}]{peruani}%
  \BibitemOpen
  \bibfield  {author} {\bibinfo {author} {\bibfnamefont {F.}~\bibnamefont
  {Peruani}}\ and\ \bibinfo {author} {\bibfnamefont {M.}~\bibnamefont
  {B{\"a}r}},\ }\href {http://stacks.iop.org/1367-2630/15/i=6/a=065009}
  {\bibfield  {journal} {\bibinfo  {journal} {New J. Phys.}\ }\textbf {\bibinfo
  {volume} {15}},\ \bibinfo {pages} {065009} (\bibinfo {year}
  {2013})}\BibitemShut {NoStop}%
\bibitem [{\citenamefont {Cremer}\ and\ \citenamefont
  {L\"owen}(2014)}]{cremer}%
  \BibitemOpen
  \bibfield  {author} {\bibinfo {author} {\bibfnamefont {P.}~\bibnamefont
  {Cremer}}\ and\ \bibinfo {author} {\bibfnamefont {H.}~\bibnamefont
  {L\"owen}},\ }\href {\doibase 10.1103/PhysRevE.89.022307} {\bibfield
  {journal} {\bibinfo  {journal} {Phys. Rev. E}\ }\textbf {\bibinfo {volume}
  {89}},\ \bibinfo {pages} {022307} (\bibinfo {year} {2014})}\BibitemShut
  {NoStop}%
\bibitem [{\citenamefont {Redner}, \citenamefont {Baskaran},\ and\
  \citenamefont {Hagan}(2013)}]{redner2013}%
  \BibitemOpen
  \bibfield  {author} {\bibinfo {author} {\bibfnamefont {G.~S.}\ \bibnamefont
  {Redner}}, \bibinfo {author} {\bibfnamefont {A.}~\bibnamefont {Baskaran}}, \
  and\ \bibinfo {author} {\bibfnamefont {M.~F.}\ \bibnamefont {Hagan}},\ }\href
  {\doibase 10.1103/PhysRevE.88.012305} {\bibfield  {journal} {\bibinfo
  {journal} {Phys. Rev. E}\ }\textbf {\bibinfo {volume} {88}},\ \bibinfo
  {pages} {012305} (\bibinfo {year} {2013})}\BibitemShut {NoStop}%
\bibitem [{\citenamefont {Schwarz-Linek}\ \emph {et~al.}(2012)\citenamefont
  {Schwarz-Linek}, \citenamefont {Valeriani}, \citenamefont {Cacciuto},
  \citenamefont {Cates}, \citenamefont {Marenduzzo}, \citenamefont {Morozov},\
  and\ \citenamefont {Poon}}]{poonpnas}%
  \BibitemOpen
  \bibfield  {author} {\bibinfo {author} {\bibfnamefont {J.}~\bibnamefont
  {Schwarz-Linek}}, \bibinfo {author} {\bibfnamefont {C.}~\bibnamefont
  {Valeriani}}, \bibinfo {author} {\bibfnamefont {A.}~\bibnamefont {Cacciuto}},
  \bibinfo {author} {\bibfnamefont {M.~E.}\ \bibnamefont {Cates}}, \bibinfo
  {author} {\bibfnamefont {D.}~\bibnamefont {Marenduzzo}}, \bibinfo {author}
  {\bibfnamefont {A.~N.}\ \bibnamefont {Morozov}}, \ and\ \bibinfo {author}
  {\bibfnamefont {W.~C.~K.}\ \bibnamefont {Poon}},\ }\href {\doibase
  10.1073/pnas.1116334109} {\bibfield  {journal} {\bibinfo  {journal} {Proc.
  Natl. Acad. Sci. USA}\ }\textbf {\bibinfo {volume} {109}},\ \bibinfo {pages}
  {4052} (\bibinfo {year} {2012})}\BibitemShut {NoStop}%
\bibitem [{\citenamefont {Sear}\ \emph {et~al.}(1999)\citenamefont {Sear},
  \citenamefont {Chung}, \citenamefont {Markovich}, \citenamefont {Gelbart},\
  and\ \citenamefont {Heath}}]{sear}%
  \BibitemOpen
  \bibfield  {author} {\bibinfo {author} {\bibfnamefont {R.~P.}\ \bibnamefont
  {Sear}}, \bibinfo {author} {\bibfnamefont {S.-W.}\ \bibnamefont {Chung}},
  \bibinfo {author} {\bibfnamefont {G.}~\bibnamefont {Markovich}}, \bibinfo
  {author} {\bibfnamefont {W.~M.}\ \bibnamefont {Gelbart}}, \ and\ \bibinfo
  {author} {\bibfnamefont {J.~R.}\ \bibnamefont {Heath}},\ }\href {\doibase
  10.1103/PhysRevE.59.R6255} {\bibfield  {journal} {\bibinfo  {journal} {Phys.
  Rev. E}\ }\textbf {\bibinfo {volume} {59}},\ \bibinfo {pages} {R6255}
  (\bibinfo {year} {1999})}\BibitemShut {NoStop}%
\bibitem [{\citenamefont {Bialk\'e}, \citenamefont {Speck},\ and\ \citenamefont
  {L\"owen}(2012)}]{bialkeprl2012}%
  \BibitemOpen
  \bibfield  {author} {\bibinfo {author} {\bibfnamefont {J.}~\bibnamefont
  {Bialk\'e}}, \bibinfo {author} {\bibfnamefont {T.}~\bibnamefont {Speck}}, \
  and\ \bibinfo {author} {\bibfnamefont {H.}~\bibnamefont {L\"owen}},\ }\href
  {\doibase 10.1103/PhysRevLett.108.168301} {\bibfield  {journal} {\bibinfo
  {journal} {Phys. Rev. Lett.}\ }\textbf {\bibinfo {volume} {108}},\ \bibinfo
  {pages} {168301} (\bibinfo {year} {2012})}\BibitemShut {NoStop}%
\bibitem [{\citenamefont {Szamel}(2014)}]{szamel}%
  \BibitemOpen
  \bibfield  {author} {\bibinfo {author} {\bibfnamefont {G.}~\bibnamefont
  {Szamel}},\ }\href {\doibase 10.1103/PhysRevE.90.012111} {\bibfield
  {journal} {\bibinfo  {journal} {Phys. Rev. E}\ }\textbf {\bibinfo {volume}
  {90}},\ \bibinfo {pages} {012111} (\bibinfo {year} {2014})}\BibitemShut
  {NoStop}%
\bibitem [{\citenamefont {Farage}, \citenamefont {Krinninger},\ and\
  \citenamefont {Brader}(2015)}]{brader}%
  \BibitemOpen
  \bibfield  {author} {\bibinfo {author} {\bibfnamefont {T.~F.~F.}\
  \bibnamefont {Farage}}, \bibinfo {author} {\bibfnamefont {P.}~\bibnamefont
  {Krinninger}}, \ and\ \bibinfo {author} {\bibfnamefont {J.~M.}\ \bibnamefont
  {Brader}},\ }\href {\doibase 10.1103/PhysRevE.91.042310} {\bibfield
  {journal} {\bibinfo  {journal} {Phys. Rev. E}\ }\textbf {\bibinfo {volume}
  {91}},\ \bibinfo {pages} {042310} (\bibinfo {year} {2015})}\BibitemShut
  {NoStop}%
\bibitem [{\citenamefont {Majumdar}, \citenamefont {Krishnamurthy},\ and\
  \citenamefont {Barma}(1998)}]{majumdar}%
  \BibitemOpen
  \bibfield  {author} {\bibinfo {author} {\bibfnamefont {S.~N.}\ \bibnamefont
  {Majumdar}}, \bibinfo {author} {\bibfnamefont {S.}~\bibnamefont
  {Krishnamurthy}}, \ and\ \bibinfo {author} {\bibfnamefont {M.}~\bibnamefont
  {Barma}},\ }\href@noop {} {\bibfield  {journal} {\bibinfo  {journal} {Phys.
  Rev. Lett.}\ }\textbf {\bibinfo {volume} {81}},\ \bibinfo {pages} {3691}
  (\bibinfo {year} {1998})}\BibitemShut {NoStop}%
\bibitem [{\citenamefont {Gregoire}\ and\ \citenamefont
  {Chat\'e}(2004)}]{chate}%
  \BibitemOpen
  \bibfield  {author} {\bibinfo {author} {\bibfnamefont {G.}~\bibnamefont
  {Gregoire}}\ and\ \bibinfo {author} {\bibfnamefont {H.}~\bibnamefont
  {Chat\'e}},\ }\href@noop {} {\bibfield  {journal} {\bibinfo  {journal} {Phys.
  Rev. Lett.}\ }\textbf {\bibinfo {volume} {92}},\ \bibinfo {pages} {025702}
  (\bibinfo {year} {2004})}\BibitemShut {NoStop}%
\bibitem [{\citenamefont {Toner}, \citenamefont {Tu},\ and\ \citenamefont
  {Ramaswamy}(2005)}]{toner}%
  \BibitemOpen
  \bibfield  {author} {\bibinfo {author} {\bibfnamefont {J.}~\bibnamefont
  {Toner}}, \bibinfo {author} {\bibfnamefont {Y.}~\bibnamefont {Tu}}, \ and\
  \bibinfo {author} {\bibfnamefont {S.}~\bibnamefont {Ramaswamy}},\ }\href@noop
  {} {\bibfield  {journal} {\bibinfo  {journal} {Annals of Physics}\ }\textbf
  {\bibinfo {volume} {318}},\ \bibinfo {pages} {170 } (\bibinfo {year}
  {2005})}\BibitemShut {NoStop}%
\bibitem [{\citenamefont {Menzel}\ and\ \citenamefont
  {L\"owen}(2013)}]{menzel}%
  \BibitemOpen
  \bibfield  {author} {\bibinfo {author} {\bibfnamefont {A.~M.}\ \bibnamefont
  {Menzel}}\ and\ \bibinfo {author} {\bibfnamefont {H.}~\bibnamefont
  {L\"owen}},\ }\href {\doibase 10.1103/PhysRevLett.110.055702} {\bibfield
  {journal} {\bibinfo  {journal} {Phys. Rev. Lett.}\ }\textbf {\bibinfo
  {volume} {110}},\ \bibinfo {pages} {055702} (\bibinfo {year}
  {2013})}\BibitemShut {NoStop}%
\bibitem [{\citenamefont {Jiang}\ and\ \citenamefont {Wu}(2009)}]{wu}%
  \BibitemOpen
  \bibfield  {author} {\bibinfo {author} {\bibfnamefont {T.}~\bibnamefont
  {Jiang}}\ and\ \bibinfo {author} {\bibfnamefont {J.}~\bibnamefont {Wu}},\
  }\href {\doibase 10.1103/PhysRevE.80.021401} {\bibfield  {journal} {\bibinfo
  {journal} {Phys. Rev. E}\ }\textbf {\bibinfo {volume} {80}},\ \bibinfo
  {pages} {021401} (\bibinfo {year} {2009})}\BibitemShut {NoStop}%
\bibitem [{\citenamefont {Z\"ottl}\ and\ \citenamefont {Stark}(2014)}]{stark}%
  \BibitemOpen
  \bibfield  {author} {\bibinfo {author} {\bibfnamefont {A.}~\bibnamefont
  {Z\"ottl}}\ and\ \bibinfo {author} {\bibfnamefont {H.}~\bibnamefont
  {Stark}},\ }\href {\doibase 10.1103/PhysRevLett.112.118101} {\bibfield
  {journal} {\bibinfo  {journal} {Phys. Rev. Lett.}\ }\textbf {\bibinfo
  {volume} {112}},\ \bibinfo {pages} {118101} (\bibinfo {year}
  {2014})}\BibitemShut {NoStop}%
\bibitem [{\citenamefont {Baraban}\ \emph
  {et~al.}(2013{\natexlab{a}})\citenamefont {Baraban}, \citenamefont {Makarov},
  \citenamefont {Schmidt}, \citenamefont {Cuniberti}, \citenamefont
  {Leiderer},\ and\ \citenamefont {Erbe}}]{baraban1}%
  \BibitemOpen
  \bibfield  {author} {\bibinfo {author} {\bibfnamefont {L.}~\bibnamefont
  {Baraban}}, \bibinfo {author} {\bibfnamefont {D.}~\bibnamefont {Makarov}},
  \bibinfo {author} {\bibfnamefont {O.~G.}\ \bibnamefont {Schmidt}}, \bibinfo
  {author} {\bibfnamefont {G.}~\bibnamefont {Cuniberti}}, \bibinfo {author}
  {\bibfnamefont {P.}~\bibnamefont {Leiderer}}, \ and\ \bibinfo {author}
  {\bibfnamefont {A.}~\bibnamefont {Erbe}},\ }\href {\doibase
  10.1039/C2NR32662K} {\bibfield  {journal} {\bibinfo  {journal} {Nanoscale}\
  }\textbf {\bibinfo {volume} {5}},\ \bibinfo {pages} {1332} (\bibinfo {year}
  {2013}{\natexlab{a}})}\BibitemShut {NoStop}%
\bibitem [{\citenamefont {Baraban}\ \emph
  {et~al.}(2013{\natexlab{b}})\citenamefont {Baraban}, \citenamefont
  {Streubel}, \citenamefont {Makarov}, \citenamefont {Han}, \citenamefont
  {Karnaushenko}, \citenamefont {Schmidt},\ and\ \citenamefont
  {Cuniberti}}]{baraban2}%
  \BibitemOpen
  \bibfield  {author} {\bibinfo {author} {\bibfnamefont {L.}~\bibnamefont
  {Baraban}}, \bibinfo {author} {\bibfnamefont {R.}~\bibnamefont {Streubel}},
  \bibinfo {author} {\bibfnamefont {D.}~\bibnamefont {Makarov}}, \bibinfo
  {author} {\bibfnamefont {L.}~\bibnamefont {Han}}, \bibinfo {author}
  {\bibfnamefont {D.}~\bibnamefont {Karnaushenko}}, \bibinfo {author}
  {\bibfnamefont {O.~G.}\ \bibnamefont {Schmidt}}, \ and\ \bibinfo {author}
  {\bibfnamefont {G.}~\bibnamefont {Cuniberti}},\ }\href {\doibase
  10.1021/nn305726m} {\bibfield  {journal} {\bibinfo  {journal} {ACS Nano}\
  }\textbf {\bibinfo {volume} {7}},\ \bibinfo {pages} {1360} (\bibinfo {year}
  {2013}{\natexlab{b}})}\BibitemShut {NoStop}%
\end{thebibliography}%
\end{document}